\documentclass{article}

\def\be{\begin{equation}}
\def\ee{\end{equation}}
\def\bea{\begin{eqnarray}}
\def\eea{\end{eqnarray}}

\def\d{\partial}

\def\tr{{\rm tr}}
\def\nnext{\vskip 10pt\noindent}

\begin{document}

\title{The Einstein-Vlasov system}

\date{}

\author{Alan D. Rendall
\\ Max-Planck-Institut f\"ur Gravitationsphysik
\\ Am M\"uhlenberg 1, 14476 Golm, Germany
}

\maketitle

\begin{abstract}\noindent
Rigorous results on solutions of the Einstein-Vlasov system are surveyed.
After an introduction to this system of equations and the reasons for
studying it, a general discussion of various classes of solutions is given.
The emphasis is on presenting important conceptual ideas, while avoiding
entering into technical details. Topics covered include spatially homogenous
models, static solutions, spherically symmetric collapse and isotropic 
singularities. 

\end{abstract}

\section{Introduction}\label{intro}

The basic equations of general relativity are the Einstein equations
coupled to some other partial differential equations describing the
matter content of spacetime. There are many choices of matter model
which are of physical interest and Yvonne Choquet-Bruhat has published
fundamental results on the Cauchy problem for the Einstein equations coupled 
to a wide variety of matter models. One of these is collisionless matter
described by the Vlasov equation. It is the subject of these lectures.

The Vlasov equation arises in kinetic theory. It gives a statistical
description of a collection of particles. It is distinguished from
other equations of kinetic theory by the fact that there is no direct
interaction between particles. In particular, no collisions are
included in the model. Each particle is acted on only by fields which
are generated collectively by all particles together. The fields which
are taken into account depend on the physical situation being
modelled. In plasma physics, where this equation is very important,
the interaction is electromagnetic and the fields are described either
by the Maxwell equations or, in a quasi-static approximation, by the
Poisson equation \cite{swanson89}. In gravitational physics, which is the 
subject of the following, the fields are described by the Einstein equations
or, in the Newtonian approximation, by the Poisson equation. (There is
a sign difference in the Poisson equation in comparison with the
electromagnetic case due to the replacement of a repulsive by an
attractive force.) The best known applications of the Vlasov equation
to self-gravitating systems are to stellar dynamics \cite{binney87}. It 
can also be applied to cosmology.  In the first case the systems considered 
are galaxies or parts of galaxies where there is not too much dust or gas
which would require a hydrodynamical treatment. Possible applications
are to globular clusters, elliptical galaxies and the central bulge of
spiral galaxies.  The \lq particles\rq\ in all these cases are
stars. In the cosmological case they might be galaxies or even clusters of
galaxies. The fact that they are modelled as particles reflects the
fact that their internal structure is believed to be irrelevant for
the dynamics of the system as a whole. The Vlasov equation is also
used in cosmology to model non-baryonic dark matter (\cite{boerner93},
p. 323). In that case the \lq particles\rq\ are elementary particles.

These lectures are concerned not with the above physical applications 
but with some basic mathematical aspects of the Einstein-Vlasov 
system. First the definition and general mathematical properties of
this system of partial differential equations are discussed and then
the Cauchy problem for the system is formulated. The central theme
in what follows is the global Cauchy problem, where \lq global\rq\ means 
global in time. The known results on this and related problems are
surveyed and important methods used are highlighted. Further
information on kinetic theory in general relativity may be found in 
\cite{ehlers73}.

Let $(M,g_{\alpha\beta})$ be a spacetime, i.e. $M$ is a four-dimensional 
manifold and $g_{\alpha\beta}$ is a metric of Lorentz signature $(-,+,+,+)$.
Note that $g_{\alpha\beta}$ denotes a geometric object here and not the
components of the geometric object in a particular coordinate system.
In other words the indices are abstract indices. (See \cite{wald84}, 
section 2.4 for a discussion of this notation.)
It is always assumed that the metric is time-orientable, i.e. that the 
two halves of the light cone at each point of $M$ can be labelled past
and future in a way which varies continuously from point to point.
With this global direction of time, it is possible to distinguish
between future-pointing and past-pointing timelike vectors. The worldline of 
a particle of non-zero rest mass $m$ is a timelike curve in spacetime. The 
unit future-pointing tangent vector to this curve is the 4-velocity 
$v^\alpha$ of the particle. Its 4-momentum $p^\alpha$ is given by 
$mv^\alpha$. There are different variants of
the Vlasov equation depending on the assumptions made. Here it is 
assumed that all particles have the same mass $m$ but it would also
be possible to allow a continuous range of masses. When all the 
masses are equal, units can be chosen so that $m=1$ and no distinction
need be made between 4-velocity and 4-momentum. There is also the 
possibility of considering massless particles, whose wordlines are
null curves. In the case $m=1$ the possible values of the four-momentum
are precisely all future-pointing unit timelike vectors. These form
a hypersurface $P$ in the tangent bundle $TM$ called the mass shell.
The distribution function $f$, which represents the density of particles
with given spacetime position and four-momentum, is a non-negative
real-valued function on $P$. A basic postulate in general relativity is
that a free particle travels along a geodesic. Consider a future-directed
timelike geodesic parametrized by proper time. Then its tangent vector
at any time is future-pointing unit timelike. Thus this geodesic has a
natural lift to a curve on $P$, by taking its position and tangent
vector together. This defines a flow on $P$. Denote the vector field
which generates this flow by $X$. (This vector field is what is sometimes
called the geodesic spray in the mathematics literature.) The condition
that $f$ represents the distribution of a collection of particles moving
freely in the given spacetime is that it should be constant along the
flow, i.e. that $Xf=0$. This equation is the Vlasov equation, sometimes
also known as the Liouville or collisionless Boltzmann equation.

To get an explicit expression for the Vlasov equation, it is necessary
to introduce local coordinates on the mass shell. In the following 
local coordinates $x^\alpha$ on spacetime are always chosen such that
the hypersurfaces $x^0$=const. are spacelike. (Greek and Latin indices take 
the values $0,1,2,3$ and $1,2,3$ respectively.) Intuitively this means that 
$x^0$, which may also be denoted by $t$, is a time coordinate and that the 
$x^a$ are spatial coordinates. A timelike vector is future-pointing if and 
only if its zero component in a coordinate system of this type is positive.
It is not assumed that the vector $\d /\d x^0$ is timelike. One way of 
defining local coordinates on $P$ is to take the spacetime coordinates 
$x^\alpha$ together with the spatial components $p^a$ of the four-momentum 
in these coordinates. Then the explicit form of the Vlasov equation is:
\be\label{vlasov}
\d f/\d t+(p^a/p^0)\d f/\d x^a-(\Gamma^a_{\beta\gamma}
p^\beta p^\gamma/p^0)\d f/\d p^a=0
\ee
where $\Gamma^\alpha_{\beta\gamma}$ are the Christoffel symbols
associated to the metric $g_{\alpha\beta}$.  Here it is understood
that $p^0$ is to be expressed in terms of $p^a$ and the metric using
the relation $g_{\alpha\beta}p^\alpha p^\beta=-1$. An alternative form of 
the Vlasov equation which is often useful is obtained by coordinatizing 
the mass shell using the components of the four-momentum in an orthonormal
frame instead of the coordinate components.

The Vlasov equation can be coupled to the Einstein equations as follows,
giving rise to the Einstein-Vlasov system. The unknowns are a 4-manifold
$M$, a (time orientable) Lorentz metric $g_{\alpha\beta}$ on $M$ and a 
non-negative real-valued function $f$ on the mass shell defined by 
$g_{\alpha\beta}$. The field equations consist of the Vlasov equation 
defined by the metric $g_{\alpha\beta}$ for
$f$ and the Einstein equation $G_{\alpha\beta}=8\pi T_{\alpha\beta}$.
(Units are chosen here so that the speed of light and the gravitational
constant both have the numerical value unity.)
To obtain a complete system of equations it remains to define $T_{\alpha
\beta}$ in terms of $f$ and $g_{\alpha\beta}$. It is defined as an integral 
over the part of the mass shell over a given spacetime point with respect to 
a measure which will now be defined. The metric at a given point of 
spacetime defines in a tautological way a metric on the tangent space
at that point. The part of the mass shell over that point is a 
submanifold of the tangent space and as such has an induced metric, 
which is Riemannian. The associated measure is the one which we are
seeking. It is evidently invariant under Lorentz transformations of
the tangent space, a fact which may be used to simplify computations in 
concrete situations. In the coordinates $(x^\alpha,p^a)$ on $P$ the
explicit form of the energy-momentum tensor is:
\be\label{em}
T_{\alpha\beta}=-\int fp_\alpha p_\beta |g|^{1/2}/p_0 dp^1 dp^2 dp^3
\ee
A simple computation in normal coordinates based at a given point shows
that $T_{\alpha\beta}$ defined by (\ref{em}) is divergence-free, independently
of the Einstein equations being satisfied. This is of course a necessary
compatibility condition in order for the Einstein-Vlasov system to be
a reasonable set of equations. Another important quantity is the
particle current density, defined by:
\be\label{current}
N^\alpha=-\int fp^\alpha |g|^{1/2}/p_0 dp^1 dp^2 dp^3
\ee
A computation in normal coordinates shows that $\nabla_\alpha N^\alpha=0$.
This equation is an expression of the conservation of the number of
particles. There are some inequalities which follow immediately from
the definitions (\ref{em}) and (\ref{current}). Firstly 
$N_\alpha V^\alpha\le 0$ for any
future-pointing timelike or null vector $V^\alpha$, with equality only if 
$f=0$ at the given point. Hence unless there are no particles at some point, 
the vector $N^\alpha$ is future-pointing timelike. Next, if $V^\alpha$ and
$W^\alpha$ are any two future-pointing timelike vectors then
$T_{\alpha\beta}V^\alpha W^\beta\ge 0$. This is the dominant energy
condition (\cite{hawking73}, p. 91). Finally, if $X^\alpha$ is a spacelike 
vector then $T_{\alpha\beta}X^\alpha X^\beta\ge 0$. This is the non-negative
pressures condition. This condition, the dominant energy condition and
the Einstein equations together imply that the Ricci tensor satisfies the
inequality $R_{\alpha\beta}V^\alpha V^\beta\ge 0$ for any timelike vector 
$V^\alpha$. The last inequality is called the strong energy condition.
These inequalities constitute one of the reasons
which mean that the Vlasov equation defines a well-behaved matter model
in general relativity. However this is not the only reason. A perfect
fluid with a reasonable equation of state or matter described by the 
Boltzmann equation also have energy-momentum tensors which satisfy these 
inequalities.

The Vlasov equation in a fixed spacetime is a linear hyperbolic equation
for a scalar function and hence solving it is equivalent to solving
the equations for its characteristics. In coordinate components 
these are:
\be\label{char}
dX^a/ds=P^a,\ \ \ dP^a/ds=-\Gamma^a_{\beta\gamma}P^\beta P^\gamma
\ee
Let $X^a(s,x^\alpha,p^a)$, $P^a(s,x^\alpha,p^a)$ be the unique 
solution of (\ref{char}) with initial conditions $X^a(t,x^\alpha,p^a)=x^a$
and $P^a(t,x^\alpha,p^a)=p^a$. Then the solution of the Vlasov equation can
be written as:
\be 
f(x^\alpha,p^a)=f_0(X^a(0,x^\alpha,p^a),P^a(0,x^\alpha,p^a))
\ee
where $f_0$ is the restriction of $f$ to the hypersurface $t=0$. This
function $f_0$ serves as initial datum for the Vlasov equation.
It follows immediately from this that if $f_0$ is bounded by some constant
$C$, the same is true of $f$. This obvious but important property of the 
solutions of the Vlasov equation is used frequently without comment in the 
study of this equation.

The above calculations involving $T_{\alpha\beta}$ and $N^\alpha$ were only
formal. In order that they have a precise meaning it is necessary to
impose some fall-off in the momentum variables on $f$ so that the integrals
occurring exist. The simplest condition to impose is that $f$ has compact
support for each fixed $t$. This property holds if the initial datum $f_0$ 
has compact support and if each hypersurface $t=t_0$ is a Cauchy hypersurface.
For by the definition of a Cauchy hypersurface, each timelike curve which 
starts at $t=0$ hits the hypersurface $t=t_0$ at a unique point. Hence the
geodesic flow defines a continuous mapping from the part of the mass shell
over the initial hypersurface $t=0$ to the part over the hypersurface $t=t_0$.
The support of $f(t_0)$, the restriction of $f$ to the hypersurface $t=t_0$
is the image of the support of $f_0$ under this continuous mapping and so
is compact. Let $P(t)$ be the supremum of the values of $|p^a|$ attained on
the support of $f(t)$. It turns out that in many cases controlling the
solution of the Vlasov equation coupled to some field equation in the
case of compactly supported initial data for the distribution function can be 
reduced to obtaining a bound for $P(t)$. An example of this is given below.

The data in the Cauchy problem for the Einstein equations coupled to any
matter source consist of the induced metric $g_{ab}$ on the initial 
hypersurface, the second fundamental form $k_{ab}$ of this hypersurface and 
some matter data. In fact these objects should be thought of as objects on
an abstract 3-dimensional manifold $S$. Thus the data consist of a 
Riemannian metric $g_{ab}$, a symmetric tensor $k_{ab}$ and appropriate
matter data, all defined intrinsically on $S$. The nature of the initial
data for the matter will now be examined in the case of the Einstein-Vlasov
system. It is not quite obvious what to do, since the distribution function
$f$ is defined on the mass shell and so the obvious choice of initial data,
namely the restriction of $f$ to the initial hypersurface, is not 
appropriate. For it is defined on the part of the mass shell over the
initial hypersurface and this is not intrinsic to $S$. This difficulty
can be overcome as follows. Let $\phi$ be the mapping which sends a 
point of the mass shell over the initial hypersurface to its orthogonal
projection onto the tangent space to the initial hypersurface. The map
$\phi$ is a diffeomorphism. The abstract initial datum $f_0$ for $f$ is
taken to be a function on the tangent bundle of $S$. The initial
condition imposed is that the restriction of $f$ to the part of the
mass shell over the initial hypersurface should be equal to $f_0$ composed
with $\phi$. An initial data set for the Einstein equations must satisfy
the constraints and in order that the definition of an abstract initial
data set for the Einstein equations be adequate it is necessary that the
constraints be expressible purely in terms of the abstract initial data.
The constraint equations are:
\bea
R-k_{ab}k^{ab}+(\tr k)^2&=16\pi\rho       \\
\nabla_a k^a_b-\nabla_b(\tr k)&=8\pi j_b
\eea
Here $R$ denotes the scalar curvature of the metric $g_{ab}$. If $n^\alpha$
denotes the future-pointing unit normal vector to the initial hypersurface
and $h^{\alpha\beta}=g^{\alpha\beta}+n^\alpha n^\beta$ is the orthogonal
projection onto the tangent space to the initial hypersurface then 
$\rho=T_{\alpha\beta}n^\alpha n^\beta$ and $j^\alpha=-h^{\alpha\beta}
T_{\beta\gamma}n^\gamma$. The vector $j^\alpha$ satisfies 
$j^\alpha n_\alpha=0$ and so can be naturally identified with a vector
intrinsic to the initial hypersurface, denoted here by $j^a$. What needs
to be done is to express $\rho$ and $j_a$ in terms of the intrinsic initial
data. They are given by the following expressions:
\bea
\rho&=\int f_0(p^a)p^a p_a/(1+p^a p_a)^{1/2}({}^{(3)}g)^{1/2} 
dp^1 dp^2 dp^3  \\
j_a&=\int f_0(p^a)p_a ({}^{(3)}g)^{1/2} dp^1 dp^2 dp^3
\eea
If a three-dimensional manifold on which an initial data set for the
Einstein-Vlasov system is defined is mapped into a spacetime by an
embedding $\psi$ then the embedding is said to induce the given initial
data on $S$ if the induced metric and second fundamental form of $\psi(S)$ 
coincide with the results of transporting $g_{ab}$ and $k_{ab}$ with $\psi$ 
and the relation $f=f_0\circ\phi$ holds, as above. A form of the local 
existence and uniqueness theorem can now be stated. This will only be done 
for the case of smooth (i.e. infinitely differentiable) initial data
although versions of the theorem exist for data of finite differentiability.

\vskip 10pt\noindent
{\bf Theorem 1.1} Let $S$ be a 3-dimensional manifold, $g_{ab}$ a smooth
Riemannian metric on $S$, $k_{ab}$ a smooth symmetric tensor on $S$
and $f_0$ a smooth non-negative function of compact support on the
tangent bundle $TS$ of $S$. Suppose further that these objects satisfy
the constraint equations. Then there exists a smooth spacetime 
$(M,g_{\alpha\beta})$, a smooth distribution function $f$ on the 
mass shell of this spacetime and a smooth embedding $\psi$ of $S$ into $M$ 
which induces the given initial data on $S$ such that $g_{\alpha\beta}$
and $f$ satisfy the Einstein-Vlasov system and $\psi(S)$ is a Cauchy 
hypersurface. Moreover, given any other spacetime $(M',g'_{\alpha\beta})$, 
distribution function $f'$ and embedding $\psi'$ satisfying these conditions, 
there exists a diffeomorphism $\chi$ from an open neighbourhood of 
$\psi(S)$ in $M$ to an open neighbourhood of $\psi'(S)$ in $M'$ which 
satisfies $\chi\circ\psi=\psi'$ and carries $g_{\alpha\beta}$ and $f$ to 
$g'_{\alpha\beta}$ and $f'$ respectively.

\vskip 10pt\noindent
The formal statement of this theorem is rather complicated, but its 
essential meaning is as follows. Given an initial data set (satisfying
the constraints) there exists a corresponding solution of the Einstein-Vlasov 
system and this solution is locally unique up to diffeomorphism. There also 
exists a global uniqueness statement which uses the notion of the maximal 
Cauchy development of an initial data set, but this is not required in 
the following.  The first proof of a theorem of the above kind for the 
Einstein-Vlasov system was given by Yvonne Choquet-Bruhat in 
\cite{choquet71}.

The problem of extending this local theorem to one which is in some sense
global is a very difficult one. In fact with presently available mathematical 
techniques it is too difficult. One way of making some progress in 
understanding the general problem is to study the simplified cases obtained 
by imposing various symmetries on the solutions. Note that if a symmetry is
imposed on the initial data for the Cauchy problem this is inherited by the
corresponding solutions. (See \cite{friedrich00}, section 5.6 for a discussion
of this.) This ensures the consistency of restricting the problem to a
particular symmetry class.

In the following different symmetry classes will be considered in turn,
proceeding from the strongest to the weakest assumptions. First spatially
homogeneous solutions are discussed. These are simple enough that it is
possible to make statements about the Einstein equations coupled to a
general class of matter models. After this has been done, further results
which can be obtained in the particular case where the matter is described
by the Vlasov equation are presented. The first inhomogeneous solutions
to be discussed are those which are static and spherically symmetric.
Apart from their intrinsic interest these are of relevance for the study
of spherically symmetric collapse which is discussed next. Brief comments
are made on dynamical cosmological solutions before concentrating on one
question where there are results on solutions of the Einstein-Vlasov 
system without any symmetry assumptions being required. This concerns
the construction of solutions with an isotropic singularity.  

\section{Spatially homogeneous solutions I: general matter models}
\label{homogeneous1}

A solution of the Einstein equations coupled to some matter equations
is said to be symmetric under the action of a Lie group $G$ if $G$
acts by isometries of the metric which also leave the matter fields
invariant. A solution of the Einstein-matter equations is called 
spatially homogeneous if it is symmetric under the action of a Lie 
group whose orbits are spacelike hypersurfaces. If we think of these
as hypersurfaces of constant time then the metric only depends on
time and the Einstein equations reduce to ordinary differential
equations, an enormous simplification. The equations of motion of 
matter fields which are 
defined on spacetime also reduce to ODE's. Since the Vlasov equation
is defined on the mass shell it in general still contains derivatives
with respect to the momenta and thus remains a partial differential
equation in the spatially homogeneous case.

The spatially homogeneous spacetimes can be classified into various 
types according to the Lie group involved. The conventional
terminology is that there are nine Bianchi types I-IX and one
additional type, the Kantowski-Sachs models. The latter will not
be discussed further here. We might like to use spatially
homogeneous spacetimes as cosmological models.  Since there is in
a sense no spatial dependence we do need to worry about spatial
boundary conditions. Eventually, however, we would like to consider
more realistic cosmological models which are inhomogeneous 
perturbations of the Bianchi spacetimes, and then boundary 
conditions become important. One simple condition to pose is 
that the spacetimes involved should contain a compact Cauchy surface.
Then there is no danger of extra information coming in from infinity.
Supposing that it is desired to impose this condition of spatial
compactness the question arises if all Bianchi types are compatible
with it. Unfortunately this is not the case. In fact the only ones
which are are types I and IX. 

A larger class of spatially compact spacetimes where the Einstein
equations reduce to ODE's are the locally spatially homogeneous
ones, as defined in \cite{rendall95a}. The idea is to require that
the spacetime itself be spatially compact while its universal cover
is spatially homogeneous. For details see \cite{rendall95a}. This
allows a much bigger class of Bianchi types to be included.

The idea now is to consider solutions of the Einstein-matter equations
which are spatially compact and locally spatially homogeneous only
assuming some general conditions on the matter model. These conditions
are satisfied in the case of the Vlasov equation but also, for example, in 
the case of the Euler equation describing a perfect fluid with a 
physically reasonable equation of state. This generality is a luxury
we can only afford due to the assumption of (local) spatial homogeneity.
In general solutions of the Euler equations can be expected to form 
shocks which leads to a breakdown of the solution of the evolution
equations. In the homogeneous case this possibility does not arise.
It is the absence of shock formation which makes the Vlasov equation
particularly convenient to work with when studying inhomogeneous 
spacetimes.

The matter models to be considered will be defined in terms of some  
general properties. As usual $T^{\alpha\beta}$ denotes the 
energy-momentum tensor. When a specific matter model has been 
chosen $T^{\alpha\beta}$ will be a functional of some matter 
variables, denoted collectively by $F$, and the spacetime metric 
$g_{\alpha\beta}$. In the following another quantity $N^\alpha$ 
(called the particle current density) will be required. It is  
also assumed to be a functional of $F$ and $g_{\alpha\beta}$. 
Now various properties which will be assumed at appropriate 
points will be listed. 
 
\nnext 
(1) $T^{\alpha\beta}V_\alpha W_\beta\ge 0$ for all future-pointing 
timelike vectors $V^\alpha$ and $W^\alpha$ (dominant energy
condition)

\nnext 
(2) $T^{\alpha\beta}(g_{\alpha\beta}+V_\alpha V_\beta)\ge 0$ for 
all unit timelike vectors $V^\alpha$ (non-negative sum pressures 
condition) 
 
\nnext 
(3) for any $F$ and $g_{\alpha\beta}$ the conditions $\nabla_\alpha 
N^\alpha=0$ and $\nabla_\alpha T^{\alpha\beta}=0$ are satisfied  
(conservation conditions) 
 
\nnext 
(4) for any $F$ and $g_{\alpha\beta}$ the vector $N^\alpha$ is  
future-pointing timelike or zero 
 
\nnext 
(5) for any constant $C_1>0$ there exists a positive constant $C_2$ 
such that for any $F$ and $g_{\alpha\beta}$ with 
$-N_\alpha N^\alpha\le C_1$ and any timelike vector $V^\alpha$ 
the following inequality holds: 
$$T^{\alpha\beta}V_\alpha V_\beta\ge C_2(N^\alpha V_\alpha)^2 $$ 
 
\nnext 
(6) for any constant $C_1>0$ there exists a positive constant $C_2<1$ 
such that for any $F$ and $g_{\alpha\beta}$ with 
$-N^\alpha N_\alpha\le C_1$ and any unit timelike vector $V^\alpha$  
$$(g_{\alpha\beta}+V_\alpha V_\beta)T^{\alpha\beta}\le 
3C_2 T^{\alpha\beta}V_\alpha V_\beta$$ 
 
\nnext 
(7) if a solution with the given symmetry of the Einstein equations 
coupled to the given matter model is such that the time 
coordinate defined above takes all values in the interval  
$(t_1,t_2)$ , if it is not possible to extend the spacetime so 
as to make this interval longer and if $t_1$ or $t_2$ is finite 
then $\tr k(t)$ is unbounded in a neighbourhood of $t_1$ or $t_2$ 
respectively.  
 
\nnext 
(8) for any constant $C_1>0$ there exists a constant $C_2>0$ such 
that $T_{\alpha\beta}T^{\alpha\beta}\le C_1$ implies $-N_\alpha
N^\alpha\le C_2$

Some comments will now be made concerning the physical motivation 
of some of these conditions. If a given type of matter 
can be considered as being made up of particles then a particle 
current density $N^\alpha$ is defined. If the particles have 
positive rest mass then this vector is future pointing timelike 
or zero as required by condition (4). If the particles are 
massless then this condition is still satisfied except 
for very special types of matter where $N^\alpha$ might be null. If 
particles cannot be created or destroyed then $N^\alpha$ is 
divergenceless as required in condition (3). It is not easy to
give an intuitive interpretation of conditions (5) and (6). The
meaning of (5) is roughly as follows. If matter is observed
from a boosted frame then the particle density is multiplied
by a $\gamma$-factor arising from the effect of Lorentz contraction
on the volume element. The observed energy density is also affected in
this way but picks up an additional $\gamma$-factor. Hence when a
given matter distribution is considered from a boosted frame the
multiplicative factor in the observed energy density behaves like
the square of that in the observed particle density. As for condition
(6), for a perfect fluid it is
related to the condition that the speed of sound should be bounded
away from the speed of light. The given symmetry
referred to in condition (7) will be the Bianchi symmetry 
being considered.

In a spatially compact spacetime an interesting quantity is the volume
of the hypersurfaces of constant time. The Friedman-Robertson-Walker
(FRW) models generally used in cosmology have the property that this
volume either increases at all times (open models) or increases to
maximum after which it decreases again (closed models). The possibility 
that it is always decreases, a case which is allowed mathematically,
is usually ignored since we know that our universe is expanding at the 
present time. In any case these models can be obtained from those where
the volume is always increasing by reversing the direction of time. Thus
they present no essentially new mathematical phenomena.

In local spatially homogeneous spatially compact spacetimes with reasonable
matter the same pattern is found. It can be proved that they also share
some other significant physical properties with the FRW models. These
spacetimes can be parametrized by a Gaussian time coordinate based on 
a (locally) homogeneous hypersurface. Suppose that the solution is maximal 
in the sense that it cannot be extended to a larger interval of Gaussian
time. Then we would like to know two things. Firstly, if the volume is
always increasing then the time of existence in the future is infinite
and the spacetime is future geodesically complete. Secondly, if the
volume is increasing at some time then the time of existence in the past
is finite and as the time of breakdown is approached some geometrical
invariant of the spacetime geometry diverges. This rules out the
possibility of extending the spacetime in some way which is not globally
hyperbolic. This is desirable form the point of view of the strong cosmic
censorship hypothesis.

In \cite{rendall95a} theorems of the desired type were proved. The first 
says that if conditions (1), (2) and (7) above are satisfied by some
matter model then inextendible locally spatially homogeneous spatially 
compact solutions of the Einstein-matter equations where the volume
is always increasing satisfy the first conclusion mentioned above. In
particular, they are future geodesically complete. The second says
that if (1)-(8) are satisfied and the spacetime is not vacuum then the 
curvature invariant $G^{\alpha\beta}G_{\alpha\beta}$ diverges as the
finite past limit of the domain of existence is approached. The 
fundamental intuitive reason for this is that a finite amount of matter
is being squeezed to zero volume as the singularity is approached.

These theorems apply to the Einstein-Vlasov system since in that case,
as will now be discussed, conditions (1)-(8) are satisfied. In that case
$T^{\alpha\beta}$ and $N^\alpha$ have been defined above. It has also
been stated that (1) and (3) hold. Condition (2) is a consequence of
the non-negative pressures condition mentioned above. To
check condition (4) it is merely necessary to observe that if $V^\alpha$
is a future-pointing timelike vector then $N_\alpha V^\alpha< 0$ unless
$N^\alpha=0$ at a given spacetime point. Condition (7) is an existence
theorem which says, roughly speaking, that as long as the geometry does
not break down too badly, the solution of the matter equations (in this
case the Vlasov equation) cannot break down in finite time. Here the mean 
curvature $\tr k$ plays the role of a controlling quantity whose boundedness
ensures the continued existence of the solution. It remains 
to check conditions (5), (6) and (8) and to do this we may 
choose a frame whose timelike member is $V^\alpha$ in order
to do the calculation. Then the inequalities of (5) and (6) become
$\hat T^{00}\ge C_2(\hat N^0)^2$ and $\delta^{ij}\hat T_{ij}\le 3C_2
\hat T^{00}$. (The hats here indicate the use of indices associated
to an orthonormal frame.)
\bea
-N_\alpha N^\alpha&=-\left(\int f(p^a)p_\alpha/p^0 dp^1 dp^2 dp^3
                     \right)
                     \left(\int f(q^a)q^\alpha/q^0 dq^1 dq^2 dq^3
                     \right)\nonumber   \\
&=-\int\int f(p^a)f(q^a)p^\alpha q_\alpha  /(p^0q^0)dp^1 dp^2 dp^3 
dq^1 dq^2 dq^3\nonumber
\\
&\ge\int\int f(p^a)f(q^a)/(p^0q^0) dp^1 dp^2 dp^3 dq^1 dq^2 dq^3\nonumber   \\
&=\left(\int f(p^a)/p^0 dp^1 dp^2 dp^3\right)^2\nonumber
\eea
Hence
\bea
\hat N^0&=\int f(p^a) dp^1 dp^2 dp^3\nonumber                  \\
   &\le \left(\int f(p^a) p^0 dp^1 dp^2 dp^3\right)^{1/2} 
      \left(\int f(p^a)/p^0
   dp^1 dp^2 dp^3\right)^{1/2}\nonumber                    \\
   &\le (\hat T^{00})^{1/2}(-N_\alpha N^\alpha)^{1/4}\nonumber
\eea
This shows that (5) holds. It follows directly from the definitions
that the inequality of (6) holds with $C_2=1/3$ even without
restricting $N_\alpha N^\alpha$ to be bounded. Finally
\bea
T_{\alpha\beta}T^{\alpha\beta}&=\left(\int f(p^a)p_\alpha p_\beta/p^0
dp^1 dp^2 dp^3\right)\left(\int f(q^a)q^\alpha q^\beta/q^0 
dq^1 dq^2 dq^3\right)\nonumber             \\
&=\int\int f(p^a)f(q^a)(p^\alpha q_\alpha)^2/(p^0q^0) dp^1 dp^2 dp^3
dq^1 dq^2 dq^3\nonumber                   \\
&\ge -\int\int f(p^a)f(q^a)(p^\alpha q_\alpha)/(p^0q^0) dp^1 dp^2 dp^3
dq^1 dq^2 dq^3\nonumber                   \\
&=-N_\alpha N^\alpha\nonumber
\eea
Thus the general theorems apply to give results on the dynamics of 
locally spatially homogeneous solutions of the Einstein-Vlasov system.
It is known that Bianchi type IX solutions cannot expand forever,
while models of the other types force the volume to be monotone.
Thus we can say the following about inextendible non-vacuum spatially
compact solutions 
of the Einstein-Vlasov system with Bianchi symmetry. If the Bianchi type 
is IX they have curvature singularities after finite proper time both in 
the past and in the future. For all other Bianchi types models which are
expanding at some time have a curvature singularity at a finite time in
the past and are future geodesically complete. The statement about 
geodesic completeness also holds in the vacuum case. The statement about
curvature singularities, however, does not. In some cases there is a Cauchy
horizon. This has been discussed in detail in \cite{chrusciel95}. See also
\cite{ringstrom00}.

\section{Spatially homogeneous solutions II: application of dynamical 
systems}\label{homogeneous2}

We have now obtained a crude picture of the dynamics of spatially
homogeneous cosmological models of the Einstein-matter equations for
a variety of matter models. It is reasonable to hope that in the case
of the Vlasov equation this can be considerably refined, in order
to get a detailed picture of the asymptotics of the models near an
initial singularity or in a phase of unlimited expansion. This has 
not yet been achieved in general but for certain cases results were
obtained in \cite{rendall00a}. These extended theorems concerning
the simpler case of massless particles obtained in \cite{rendall99a}.
The models considered were of the simplest Bianchi types I, II and III.
It was assumed that a further symmetry is present so that the 
spacetimes have a total of four Killing vectors. These are the so-called
LRS models (locally rotationally symmetric). The reason for 
making this assumption is that then the Vlasov equation can be solved 
explicitly and only the Einstein equations remain to be handled. The
equations can be reduced to a system of ODE's in contrast to the general
case, although the coefficients of the system involve one function
which is not known explicitly. Fortunately it suffices to know certain
qualitative features of this function in order to determine the
asymptotic behaviour of the solutions of the ODE's at early and late 
times. In fact to do this analysis it was also necessary to assume
invariance under certain reflections, a piece of information which 
will be suppressed in the following for simplicity.

To describe the results it is useful to introduce the generalized Kasner
exponents $p_i$. Each of the homogeneous hypersurfaces has an induced
metric $g_{ab}$ and a second fundamental form $k_{ab}$. Let $\lambda_i$ 
denote the
eigenvalues of the second fundamental form with respect to the metric.
By definition this means that they are the solutions of the eigenvalue
equation $\det (k_{ab}-\lambda g_{ab})=0$. Suppose now that $\tr k$ is
non-zero. Then we can define $p_i=\lambda_i/(\sum_j \lambda_j)$. The
quantities $p_i$ are functions of $t$ and their sum is equal to one.
The Kasner solution of the vacuum Einstein equations is given by
\be
ds^2=-dt^2+t^{2p_1}dx^2+t^{2p_2}dy^2+t^{2p_3}dz^2
\ee
for the Kasner exponents $p_i$, which are constants satisfying
$\sum_j p_j=1$ and $\sum_j p_j^2=1$. These equations are called
the first and second Kasner relations. In this case the 
generalized Kasner exponents are equal to the quantities
$p_i$ in this metric form and this explains their name.

Consider a solution of the Einstein-Vlasov system of Bianchi type I
which is LRS. Choose the time of the initial singularity to be $t=0$.
The following statements hold. For each $i$ the quantity $p_i(t)$
converges to $1/3$ as $t\to\infty$. This is the value of the 
generalized Kasner exponents in a spatially flat FRW model. This
means that the spacetime isotropizes at late times. At early
times there are three possibilities. The first is that the $p_i$
are identically equal to $1/3$ at all times. This is the FRW case.
The second is that they have limits $(1/2,1/2,0)$. The third,
which is the generic case, is that they have the limits 
$(2/3,2/3,-1/3)$. In this last case the second Kasner relation is
satisfied asymptotically and so, in a certain sense, the solution
with matter is approximated near the singularity by a vacuum 
solution.

In type III the initial singularity is similar to that in type I
but in type II it is oscillatory. As $t\to 0$ the generalized
Kasner exponents approach both the values $(2/3,2/3,-1/3)$ and 
$(0,0,1)$ as closely as desired in any neighbourhood of $t=0$.
In the expanding direction the type II solution approaches an 
explicitly known self-similar dust spacetime whose generalized Kasner 
exponents have the values $(3/8,3/8,1/4)$. In type III the generalized
Kasner exponents approach $(0,0,1)$ in the expanding direction.
It is possible to get more detailed information about the asymptotics
of these models in the limits $t\to 0$ and $t\to\infty$. The proofs
of all these statements make use of the fact that for ODE's the highly
developed theory of dynamical systems is available. Getting out the
finer features of the expanding phase of Bianchi III models is more
delicate that the other cases just discussed and was carried out using
centre manifold theory in \cite{rendall01a}.

A key aspect of the work on the asymptotics of types I, II and III was
writing the dynamical system in cleverly chosen variables. As we go
to a singularity or to infinity in a phase of unlimited expansion the
most obvious variables go to zero or infinity. If dimensionless variables
can be found which, at least for some of the solutions, converge to
finite non-zero limits in these regimes then this is a great help in
analysing the asymptotics. Often the original dynamical system can be
extended to a smooth dynamical system on a compact region. This avoids
the problem, found with many choices of variables, that the solutions
expressed in the given variables run off to infinity in a way which
is hard to control. The strategy of dimensionless variables and
compactification has been carried much further in the case of spatially
homogeneous solutions of the Einstein-Euler system. For an account of
this see \cite{wainwright97a}.

\section{Static solutions}\label{static}

This section is concerned with static spherically symmetric solutions of 
the Einstein-Vlasov system. These may play a role in describing the
long-time behaviour of solutions of the
full dynamical equations and so they have a natural place in these 
lectures. Before coming to the Einstein-Vlasov system it is worth spending
a little time thinking about the corresponding non-relativistic problem,
where a lot more is known. There are results on static solutions of the
Vlasov-Poisson system which are not spherically symmetric and stationary
solutions which are not static \cite{rein00a}. Nothing comparable has yet 
been done in the case of the Einstein-Vlasov system. This is a gap 
which should be filled. From now on we restrict consideration to the static
spherically symmetric case.

There are two methods which have been used to construct static spherically
symmetric solutions of the Vlasov-Poisson system. The first may be called
the ODE method. In a spherically symmetric static spacetime there are two
constants of motion of the particles, namely the energy $E$ and the modulus 
of the angular momentum $L$, which are useful in constructing solutions of
the Vlasov equation. In fact $E$ and $L$, like any quantity conserved along
geodesics, are solutions of the Vlasov equation. The same is true of any
function $\Phi(E,L)$ of these quantities. Jeans' theorem says that in a 
spherically symmetric static solution of the Vlasov-Poisson system the
distribution function is a function of $E$ and $L$. Thus a natural
procedure is to make an ansatz for the distribution function by choosing
a particular function $\Phi$. The Einstein equations then reduce to a system
of integrodifferential equations for the metric coefficients as functions
of a radial coordinate. What remains to be done is to analyse the global
properties of solutions of this system.

The second method is a variational one. The Vlasov-Poisson system can be
expressed as an infinite-dimensional Hamiltonian system. (Cf. \cite{holm85}.)
It is degenerate
in the sense that instead of a symplectic structure there is only a
Poisson structure. This leads to a large class of conserved quantities
known as Casimir invariants. In a Hamiltonian system a minimum of the 
Hamiltonian is a time-independent solution of the equations of motion.
This suggests a variational route to finding static solutions. When 
Casimir invariants are present there are more general possibilities.
If $C$ is a Casimir invariant then a minimum of $H+C$ is a time-independent
solution of the equations of motion. An advantage of this method is that
apart from giving results on the existence of static solutions it can also 
provide information on the stability of the solutions obtained. This 
method has been applied extensively to the Vlasov-Poisson system by Guo
and Rein (see \cite{rein02a} and references therein).

Returning to the Einstein-Vlasov system, there is a paper \cite{wolansky}
where the energy-Casimir method has been applied but it seems to be much
harder than the Vlasov-Poisson case and the results are much more limited.
More straightforward is the ODE method. One cautionary note is in order.
The direct analogue of Jeans' theorem is not true in the case of the 
Einstein-Vlasov system. Counterexamples were constructed by Schaeffer
\cite{schaeffer99a}.
Nevertheless we can still assume a distribution function of the form 
$\Phi(E,L)$ and procede from there. A theorem on global existence in the
radial coordinate for a rather general choice of $\Phi$ was obtained in
\cite{rein94a}. There is a difficulty concerning the physical relevance 
of these solutions. If we would like to use them to model globular clusters,
for instance, then we would like to obtain configurations of finite total
mass. The easiest way to prove this is if the spatial density has compact
support. The general existence theorem does not give any information on
this. In fact whether it is true or not depends on the choice of the 
function $\Phi$ in quite a delicate way. A criterion for the finiteness
of the mass in a large class of functions $\Phi$ was given in 
\cite{rein00b}. 

All known static solutions of the Einstein-Euler system with a physically 
reasonable equation of state are spherically symmetric and the density is
a monotone decreasing function of the radius. In the case of the 
Einstein-Vlasov system another kind of configuration is possible where
the support of the density is a thick shell, i.e. the region between
two concentric spheres. In order to achieve this the function $\phi$ must
depend on the angular momentum. If it only depends on the energy then
the system is equivalent to a solution of the Einstein-Euler system
with an equation of state which is in general not explicitly known.
The existence of shell solutions was proved in \cite{rein99a}.  

\section{Spherically symmetric collapse}\label{spherical}

An interesting situation to consider is that of an isolated system consisting
of matter undergoing gravitational collapse. The traditional model for
this, following Oppenheimer and Snyder, is the collapse of a homogeneous
spherical 
cloud of dust. Unfortunately when inhomogeneities are introduced into
the Oppenheimer-Snyder model they often lead to pathologies such as
shell-crossing singularities. The advantage of using collisionless matter
is that it avoids some of (and perhaps all) the problems associated with
dust.

A natural first step towards understanding spherical collapse is to
fully understand the case where there is no collapse. If we have only a 
small amount of matter then it is to be expected that its self-gravitation
will not suffice to keep it together and that it will spread out and 
disperse to infinity. For collisionless matter this has been proved, as
described in more detail later. For dust it is not true since even small
amounts of matter can develop shell-crossing singularities. Even dust
without gravitation can do so and so this effect has nothing to do with
gravity at all.

Consider initial data for the Einstein-Vlasov system which are spherically
symmetric and asymptotically flat. In a suitable coordinate system only 
data for the distribution function need be given since the metric can then
be determined by solving equations on the hypersurfaces of constant time.
This is a reflection of the familiar statement that there is no gravitational
radiation in spherical symmetry. Now let us make the initial data small in
the following sense. In the presence of fixed bounds on the extent of the
support of the initial data for $f$ in position and velocity space we require
the maximum of $f$ to be small. For small data it can be shown that the
solution exists globally in a suitable time coordinate and, more 
importantly, that it is geodesically complete. Moreover, various 
quantities such as the energy density of the matter decay to zero as
$t\to\infty$ \cite{rein92a}. Thus it can be seen that for small data the
solution disperses and the situation is completely under control.

What happens for large data? It is known that for large data a trapped 
surface (and presumably a black hole) can form \cite{rendall92a}. We might 
nevertheless get global existence in a singularity-avoiding time slicing 
like maximal or polar slicing. The latter is also sometimes called a 
Schwarzschild time coordinate. There is a theorem \cite{rein95a}
which says that if a singularity forms in a solution of the spherically
symmetric Einstein-Vlasov system then the first singularity (as measured
in Schwarzschild time) occurs at the centre of symmetry. Note that the
shell-crossing singularities of dust occur away from the centre. A
corresponding result for maximal slicing has been proved in 
\cite{rendall97a}.

A general mathematical result on the behaviour of spherically symmetric 
asymptotically flat solutions of the Einstein-Vlasov system has not yet 
been obtained. In the absence of further analytical progress, attempts
have been made to study the problem numerically. One theme which plays 
an important role is that of critical collapse. Suppose that we have a
family of data depending on a parameter $\lambda$ for the spherically 
symmetric Einstein-Vlasov system which interpolates between weak and 
strong data, with $\lambda=0$ corresponding to data for flat space. For
$\lambda$ sufficiently small the theorem already mentioned tells us that
the matter disperses. For $\lambda$ sufficiently large we might expect 
collapse to a black hole and this is indeed seen numerically. More 
precisely, it is seen that for $\lambda$ smaller than a certain value
$\lambda_*$ the matter disperses while for $\lambda>\lambda_*$ it 
collapses to a black hole. In general some of the matter falls into the 
black hole while some escapes. Let $M(\lambda)$ be the mass of the black
hole formed when the initial data corresponds to the parameter value
$\lambda$. If no black hole is formed $M(\lambda)$ is defined to be zero.
One of the questions which comes up in the study of critical collapse
is whether the function $M(\lambda)$ is continuous at $\lambda_*$ or 
not.

In \cite{rein98a} numerical evidence was presented that $M(\lambda)$ is
not continuous. In other words, the limit of $M(\lambda)$ as 
$\lambda\to\lambda_*$ from above is strictly positive. This is different
from what is found for some other matter models, such as the massless
scalar field. Olabarrieta and Choptuik \cite{olabarrieta} confirmed
this finding and were able to present a more detailed picture of what
happens. It is convenient for the numerical calculations to take 
initial data where there are no particles at the centre and no particles
on purely radial orbits. In that case as long as the solution remains
regular no particle can reach the centre due to conservation of angular
momentum. Thus we have a dynamical configuration of matter with a hole
in the middle. This allows difficulties with the singularity of polar
coordinates at $r=0$ to be avoided. It is found in \cite{olabarrieta}
that the solution evolves towards an unstable static shell solution
before turning away again and dispersing or collapsing. The mass of
the shell solution sets the mass gap in the graph of $M(\lambda)$. The
connection between the shell solutions observed numerically in
collapse calculations and those whose existence has been shown rigorously
is not clear.

\section{Isotropic singularities and Fuchsian methods}\label{isotropic}

In the last section results for certain asymptotically flat solutions of the
Einstein-Vlasov system were described. These are spherically symmetric and 
hence have three Killing vectors. There seem to be no other symmetry 
assumptions on asymptotically flat spacetimes which can be usefully
studied at the present time. The obvious symmetry class which comes to mind
is axial symmetry. In that case, however, there is only one Killing vector, 
which is very little, and even that has fixed points, which leads to 
singularities
of the equations obtained when the symmetry is factored out. At the moment
it seems to offer no advantage over the general case. In the case of 
spacetimes evolving from data on a compact Cauchy surface (cosmological
spacetimes) there is a variety of interesting symmetry types with two or
three Killing vectors and a number
of papers on solutions of the Einstein-Vlasov system with these symmetries.
They will not be reviewed here since some choices had to be made in order
to limit the volume of the lectures. A good point of entry into the 
literature is \cite{andreasson01a}. 
  
There is one mathematical result on the Einstein-Vlasov system which does
not require any symmetry assumptions and it will be the subject of the 
remainder of this section. It concerns solutions of the Einstein-Vlasov
system with massless particles and it would be interesting to know if 
analogous results hold for massive particles. The idea is to construct 
large classes of solutions of the equations whose singularities have a 
particular structure, the isotropic singularities.

Given a spacetime with a foliation by spacelike hypersurfaces we can define 
the generalized Kasner exponents as in section \ref{homogeneous2} in 
terms of the eigenvalues of the second fundamental form. In the 
inhomogeneous case these are functions $p_i$ on spacetime which in
general depend on both the time and space coordinates. The condition
for an isotropic singularity (at least intuitively) is that all
generalized Kasner exponents should tend to $1/3$ as the singularity
is approached. Thus the solution looks like an isotropic FRW model
near the singularity. The actual definition used in the theorem is
a different one.  A spatially flat FRW model is conformally flat. In
the definition of an isotropic singularity it is assumed that the given 
metric is conformal to a metric which is smooth at the singularity.

The requirement of an isotropic singularity is a restriction on the 
spacetimes considered but it has been proved by Anguige \cite{anguige00a}
that there is a very large class of solutions of the Einstein-Vlasov 
system with isotropic singularities. In particular, he does not have
to make any symmetry assumptions. The solutions can be parametrized by
certain data on the singularity which can be given freely. The method of 
proving this is to
use Fuchsian methods (although Anguige does not use this terminology).
This technique is of wider importance in studying singularities of
solutions of the Einstein-matter equations and will now be discussed 
in a more general context.

Let a system of partial differential equations with smooth coefficients
be given and suppose we would like to investigate the existence of solutions
which become singular on a certain hypersurface. For simplicity assume
that this is the coordinate hypersurface $t=0$. Suppose that in some way
it was possible to guess the asymptotic behaviour of the solutions in the
approach to the singularity. This might be done by studying explicit
solutions or by using trial and error to get a formally consistent 
asymptotic expansion. Then express the solution $u$ to be constructed
in terms of an explicit function $u_0$ having the expected asymptotics
near the singularity and a remainder $v$ which is expected to be regular
and vanish at $t=0$. Now rewrite the original equation for the unknown $u$
as an equation for $v$ whose coefficients depend on $u_0$. Since $u_0$
is singular it is to be expected that the equation for $v$ is singular
at $t=0$. Thus the problem of finding a singular solution of a regular 
equation has been replaced by that of finding a regular solution of a 
singular equation.

In favourable cases the singular equation obtained by this method is
a Fuchsian equation of the form
\be
t\d_t v+N(x)v=tf(t,x,v,v_x)
\ee  
where the matrix-valued function $N$ has some positivity property. There
are theorems which guarantee that an equation of this kind has a unique
solution $v$ which is regular and vanishes at $t=0$. One theorem of this
kind was proved in \cite{kichenassamy}, where it was applied to study
singularities in Gowdy spacetimes. Since then there have been a number 
of other applications. (See \cite{rendall02a}, section 6.2, for more
information on this.) Unfortunately the Einstein-Vlasov system does not 
fit into the framework of this theorem due to the fact that it is an
integrodifferential equation rather than a differential equation. For
this reason Anguige had to prove his theorem by doing a direct iteration.

\section{Outlook} 

In these lectures a selection of work on the Einstein-Vlasov system has 
been surveyed. Although the results were only discussed on a very general 
level, without getting into details, it was still necessary to leave out 
a lot of interesting topics. Some of them were mentioned briefly, some not 
at all. It should be clear that this is an area of research where there
are many open problems and many promising directions to be explored. The
references given here should provide a good starting point for those 
wanting to follow this road.

\end{document}